\documentclass{article}

\usepackage{PRIMEarxiv}

\usepackage[utf8]{inputenc}
\usepackage[T1]{fontenc}
\usepackage{hyperref}
\usepackage{url}
\usepackage{booktabs}
\usepackage{amsfonts}
\usepackage{amsmath}
\usepackage{microtype}
\usepackage{graphicx}
\usepackage[section]{placeins}

\usepackage{tikz}

\title{
A Compact Phenomenological Pattern in Fermion Mass Ratios and Mixing Parameters
}

\author{
Petr~Baron
}

\begin{document}

\maketitle

\begin{abstract}
We record a compact numerical regularity in charged-fermion mass ratios and
fermion mixing parameters.  The construction uses discrete generation labels
\(G=1,2,3\), fixed structural integers \(N_c=3\) and \(N_w=2\), sector
exponent functions \(L_A(G)\), and simple phase assignments.  No continuous
numerical optimization is performed: after one overall mass scale is chosen
in each charged sector, the charged-fermion mass ratios and the leading
mixing inputs follow directly from the stated formulae.  The same exponent
structure generates six charged-fermion mass ratios, six mixing sines, and
the CKM phase as correlated numerical outputs.  The full CKM and PMNS
matrices are then obtained by standard unitary reconstruction.  The
construction is phenomenological and does not claim a first-principles
derivation of the numerical constants.
\end{abstract}

\keywords{
Fermion Masses \and Flavor Physics \and Quark Mixing \and Lepton Mixing
\and CKM Matrix \and PMNS Matrix \and Mass Hierarchies
}

\section{Introduction}

The origin of the observed pattern of fermion masses and mixing angles
remains one of the open structural questions of the Standard Model.  In the
Standard Model, quark and charged-lepton masses arise from Yukawa couplings
after electroweak symmetry breaking, while quark mixing is described by the CKM matrix
\cite{Cabibbo1963,KobayashiMaskawa1973}.  Neutrino oscillation data require
lepton mixing, described by the PMNS matrix
\cite{Pontecorvo1957,MakiNakagawaSakata1962}, and at least two nonzero neutrino masses.  
The numerical values of the Yukawa couplings and mixing parameters, however, are not
predicted by the Standard Model itself; they are taken as experimental
inputs.  This situation is usually referred to as the flavor problem.

A large literature has attempted to organize or explain this pattern.  Among
the main approaches are texture relations for fermion mass matrices
\cite{Fritzsch1977,Fritzsch1978}, Froggatt--Nielsen mechanisms based on
horizontal charge assignments \cite{FroggattNielsen1979}, grand-unified mass
relations \cite{GeorgiJarlskog1979}, continuous and discrete flavor
symmetries \cite{AltarelliFeruglio2010,KingLuhn2013,DingValle2024},
modular flavor symmetries \cite{Feruglio2019,ChenRatz2025}, and
seesaw-based constructions for neutrino masses
\cite{Minkowski1977,Yanagida1979,GellMannRamondSlansky1979,MohapatraSenjanovic1980}. 
These frameworks differ in their dynamical assumptions, but they
share the goal of reducing the apparent arbitrariness of the fermion mass and
mixing parameters.

The purpose of the present work is more modest.  We record a compact
phenomenological regularity in which a common set of generation-dependent
exponents controls charged-fermion mass ratios and the leading inputs to
quark and lepton mixing.  The construction uses the generation label
\[
G=1,2,3,
\]
the fixed integers
\[
N_c=3,\qquad N_w=2,
\]
and sector-dependent exponent functions \(L_A(G)\), with
\[
A\in\{u,d,\nu,e\}.
\]
The same exponents determine the charged-fermion mass ratios and the
amplitude ratios entering the mixing formula.  After the three independent
mixing sines and the corresponding phase are obtained, the CKM and PMNS
matrices are reconstructed using the standard unitary parametrization.

The construction should not be interpreted as a complete flavor model.  No
new gauge symmetry, scalar sector, ultraviolet completion, or dynamical
mechanism is proposed.  In particular, the exponent functions and bridge
factors used below are not derived from a microscopic theory.  Instead, the
aim is to display an economical numerical structure and to separate clearly
which quantities are fixed inputs and which quantities are correlated
outputs.

The main observation is that, once one overall mass scale is chosen in each
charged sector, the leading expression generates six charged-fermion mass
ratios.  The same exponent structure also generates six mixing sines, three
for the CKM sector and three for the PMNS sector, together with the phase
used for the CKM reconstruction.  Since the mass ratios and mixing inputs
are derived from the same functions \(L_A(G)\), they should be regarded as
correlated outputs rather than independent fits.

The paper is organized as follows.  Section~2 defines the compact unified
mixing formula and the exponent functions.  Section~3 evaluates the charged
mass ratios and the CKM and PMNS matrices.  Section~4 summarizes the
numerical economy and limitations of the construction.

\section{Definition of the Construction}

We define the generation labels
\[
G=1,2,3.
\]
The fixed numerical constants are
\[
N_c=3,
\qquad
N_w=2,
\qquad
\Phi_0=\frac{\pi}{2}.
\]
The elementary angular factors are
\[
b_{60}=\cos\frac{\pi}{3},
\qquad
b_{45}^{+}
=
\cos\frac{\pi}{4}
+
\sin\frac{\pi}{4}.
\]

In order to write the CKM and PMNS cases in a single form, we introduce
sector selector symbols. For each sector \(A\in\{u,d,\nu,e\}\), define
\[
\delta_{Aq}=\delta_{Au}+\delta_{Ad},
\qquad
\delta_{A\ell}=\delta_{A\nu}+\delta_{Ae},
\]
and
\[
\sigma_A
=
\delta_{Au}+\delta_{A\nu}
-
\delta_{Ad}-\delta_{Ae}.
\]
Thus
\[
\sigma_u=\sigma_\nu=+1,
\qquad
\sigma_d=\sigma_e=-1.
\]

For the CKM sector one uses
\[
(A,B)=(d,u),
\]
and defines
\[
s^q_{ij}=|M^{(d,u)}_{ij}|.
\]
For the PMNS sector one uses
\[
(A,B)=(e,\nu),
\]
and defines
\[
s^\ell_{ij}=|M^{(e,\nu)}_{ij}|.
\]
The full CKM and PMNS matrices are then reconstructed from the three mixing sines and the corresponding CP phase.

\begin{equation}
\boxed{
|M^{(A,B)}_{ij}|
=
\mathcal G_{ij}\,
e^{-S_{ij}/2}\,
3^{|B^{(A,B)}_{ij}|^2/2}
\sqrt{
\left(R^{(A)}_{ij}\right)^2
+
\left(R^{(B)}_{ij}\right)^2
-
2R^{(A)}_{ij}R^{(B)}_{ij}
\cos\Phi^{(A,B)}_{ij}
}\, .
}
\label{eq:mixing-compact}
\end{equation}

The factors in Eq.~\eqref{eq:mixing-compact} are defined as follows.  The
$\mathcal G_{ij}$ is
\begin{equation}
\mathcal G_{ij}
=
\left(
\frac{2G_i-1}{2G_j-1}
\right)^{3/2}.
\label{eq:geometric-prefactor}
\end{equation}
The $S_{ij}$ factor is
\begin{equation}
S_{ij}
=
\left((G_j-G_i)(G_j-2)\right)^2
+
\left(b_{60}\delta_{i1}\delta_{j3}\right)^2.
\label{eq:shell-action}
\end{equation}
The $B^{(A,B)}_{ij}$ factor is
\begin{equation}
B^{(A,B)}_{ij}
=
\left(N_c^2-1\right)^{-\delta_{Ad}\delta_{Bu}}
\left[
b_{60}\delta_{i1}\delta_{j2}
+
\delta_{i2}\delta_{j3}
+
b_{45}^{+}\delta_{i1}\delta_{j3}
\right].
\label{eq:bridge-factor}
\end{equation}
The phase is taken as
\begin{equation}
\Phi^{(A,B)}_{ij}
=
\frac{\pi}{2}\,
\delta_{Ad}\delta_{Bu}\delta_{i1}\delta_{j2}.
\label{eq:phase-factor}
\end{equation}

The ratios \(R^{(A)}_{ij}\) are generated from the same exponent structure
that controls the mass hierarchy.  It is useful to write the sector
exponents in weak-doublet form.  For \(X=q,\ell\), define
\[
L_{X,\pm}(G)
=
\frac12
\left[
C_X(G)
\pm
D_X(G)
\right],
\]
where the upper sign denotes the upper member of the weak doublet and the
lower sign denotes the lower member.  The physical sector exponents are
identified as
\[
L_u=L_{q,+},
\qquad
L_d=L_{q,-},
\qquad
L_\nu=L_{\ell,+},
\qquad
L_e=L_{\ell,-}.
\]

The doublet-center functions are
\[
C_q(G)
=
(N_c+N_w)
\left(
G+\frac12\delta_{G1}
\right),
\]
and
\[
C_\ell(G)
=
\frac{N_w}{N_c}
\left(
G-\frac12\delta_{G1}
\right).
\]
The corresponding doublet splittings are
\[
D_q(G)
=
G-2\delta_{G1},
\]
and
\[
D_\ell(G)
=
\frac{(-1)^G G!}{N_c}
+
\left[
\sum_{r=1}^{2}(2r-1)
\right]
b_{45}^{+}\delta_{G1}.
\]

Equivalently, the four explicit sector exponents are
\[
L_u(G)
=
\frac12
\left[
C_q(G)+D_q(G)
\right],
\qquad
L_d(G)
=
\frac12
\left[
C_q(G)-D_q(G)
\right],
\]
and
\[
L_\nu(G)
=
\frac12
\left[
C_\ell(G)+D_\ell(G)
\right],
\qquad
L_e(G)
=
\frac12
\left[
C_\ell(G)-D_\ell(G)
\right].
\]

The amplitude ratios entering the mixing formula are then
\begin{equation}
\boxed{
R^{(A)}_{ij}
=
3^{-\frac12\left[L_A(G_j)-L_A(G_i)\right]}.
}
\label{eq:amplitude-ratio}
\end{equation}
Thus,
\[
R^{(u)}_{ij}
=
3^{-\frac12\left[L_{q,+}(G_j)-L_{q,+}(G_i)\right]},
\qquad
R^{(d)}_{ij}
=
3^{-\frac12\left[L_{q,-}(G_j)-L_{q,-}(G_i)\right]},
\]
and
\[
R^{(\nu)}_{ij}
=
3^{-\frac12\left[L_{\ell,+}(G_j)-L_{\ell,+}(G_i)\right]},
\qquad
R^{(e)}_{ij}
=
3^{-\frac12\left[L_{\ell,-}(G_j)-L_{\ell,-}(G_i)\right]}.
\]

The same exponents determine the charged-fermion mass ratios through
\begin{equation}
\boxed{
\frac{m_A(G_j)}{m_A(G_i)}
=
3^{L_A(G_j)-L_A(G_i)}
\left(
\frac{2G_j-1}{2G_i-1}
\right)^{p_A}.
}
\label{eq:mass-ratio}
\end{equation}
Here
\[
p_u=p_d=N_c,
\qquad
p_\nu=p_e=N_w.
\]
Thus Eq.~\eqref{eq:mass-ratio} shows that \(L_A(G)\) controls the
single-sector mass hierarchy, while Eq.~\eqref{eq:amplitude-ratio} gives its
square-root inverse form entering the transition amplitudes.

The same center structure gives a compact geometric representation of the
CKM phase.  We define the two coefficient-space center vectors
\[
\vec C_q=
\left(
N_c+N_w,\; +1
\right),
\qquad
\vec C_\ell=
\left(
\frac{N_w}{N_c},\; -1
\right).
\]
The first component gives the slope of the doublet-center function, while
the second component records the sign of the first-generation half-shift.
The relative angle between these two center directions is
\begin{equation}
\boxed{
\cos\alpha_{q\ell}
=
\frac{
\vec C_q\cdot\vec C_\ell
}{
|\vec C_q|\,|\vec C_\ell|
}.
}
\end{equation}
Explicitly,
\[
\vec C_q\cdot\vec C_\ell
=
(N_c+N_w)\frac{N_w}{N_c}-1,
\]
and
\[
|\vec C_q|
=
\sqrt{(N_c+N_w)^2+1},
\qquad
|\vec C_\ell|
=
\sqrt{\left(\frac{N_w}{N_c}\right)^2+1}.
\]
Therefore
\begin{equation}
\boxed{
\alpha_{q\ell}
=
\arccos\left(\frac{
(N_c+N_w)\frac{N_w}{N_c}-1
}{
\sqrt{(N_c+N_w)^2+1}
\sqrt{\left(\frac{N_w}{N_c}\right)^2+1}
}
\right)
\simeq 67.62^\circ .
\label{eq:ckm-phase-angle}
}
\end{equation}
We identify the CKM phase with the relative center orientation defined in
Eq.~\eqref{eq:ckm-phase-angle},
\[
\boxed{
\alpha_q\equiv\alpha_{q\ell}
\simeq 67.62^\circ .
}
\]

This completes the definition of the input structure.  In the following
section we evaluate the resulting charged-fermion mass ratios, CKM mixing
matrix, and PMNS mixing matrix.

\section{Numerical Results}

We now evaluate the formula using the fixed structural inputs
\[
N_c=3,\qquad N_w=2,\qquad
b_{60}=\cos\frac{\pi}{3},\qquad
b_{45}^{+}=\cos\frac{\pi}{4}+\sin\frac{\pi}{4}.
\]
No continuous numerical fit is performed in the mixing sector.  The formula
first gives the independent mixing sines \(s_{12},s_{23},s_{13}\), and the
full CKM and PMNS matrices are then reconstructed using the standard unitary
parametrization.

The reference values used for charged-fermion masses and mixing parameters
are taken from standard particle-data and global-fit sources
\cite{PDG2024,FLAG2024,NuFIT2024}.

\subsection{Charged-Fermion Mass Ratios}

The charged-fermion mass ratios are obtained from the pure leading exponent structure.  The agreement should be
interpreted as leading order; residual deviations may be represented by
small external dressing factors discussed later.

\begin{table}[h!]
\centering
\caption{Charged-fermion mass ratios from the leading exponent structure.}
\label{tab:mass-ratios}
\begin{tabular}{lrrrr}
\hline
Ratio & Model & Reference & Difference & Rel. diff. [\%]\\
\hline
\(m_c/m_u\)       & 553.920 & 587.963 & -34.043 & -5.790\\
\(m_t/m_c\)       & 125.000 & 135.882 & -10.882 & -8.008\\
\(m_s/m_d\)       & 20.516  & 20.000  & 0.516   & 2.578\\
\(m_b/m_s\)       & 41.667  & 44.754  & -3.087  & -6.898\\
\(m_\mu/m_e\)     & 201.254 & 206.768 & -5.514  & -2.667\\
\(m_\tau/m_\mu\)  & 17.334  & 16.818  & 0.516   & 3.070\\
\hline
\end{tabular}
\end{table}

\subsection{Mixing Sines}

The formula gives the following independent mixing sines:
\[
\begin{array}{c|ccc}
 & s_{12} & s_{23} & s_{13}\\
\hline
\mathrm{CKM}  & 0.225313537 & 0.040055899 & 0.003693947\\
\mathrm{PMNS} & 0.556078168 & 0.650218338 & 0.148832147
\end{array}
\]
For the CKM reconstruction we use
\[
\alpha_q
= 
67.62^\circ .
\]

For the PMNS modulus comparison shown below we use \(\alpha_\ell=0\).

\subsection{CKM Matrix}

The reconstructed CKM matrix is
\[
|V_{\rm CKM}|_{\rm model}
=
\begin{pmatrix}
0.974280 & 0.225312 & 0.003694\\
0.225188 & 0.973492 & 0.040056\\
0.008347 & 0.039350 & 0.999191
\end{pmatrix}.
\]
The comparison with the reference values is shown in
Table~\ref{tab:ckm-comparison}.

\begin{table}[h!]
\centering
\caption{Comparison of the reconstructed CKM matrix with reference values.}
\label{tab:ckm-comparison}
\begin{tabular}{crrrr}
\hline
Entry & Model & Reference & Difference & Rel. diff. [\%]\\
\hline
11 & 0.974280 & 0.974010 & 0.000270  & 0.027686\\
12 & 0.225312 & 0.226500 & -0.001188 & -0.524503\\
13 & 0.003694 & 0.003610 & 0.000084  & 2.325392\\
21 & 0.225188 & 0.226360 & -0.001172 & -0.517919\\
22 & 0.973492 & 0.973200 & 0.000292  & 0.029972\\
23 & 0.040056 & 0.040530 & -0.000474 & -1.170427\\
31 & 0.008347 & 0.008540 & -0.000193 & -2.261418\\
32 & 0.039350 & 0.039780 & -0.000430 & -1.080766\\
33 & 0.999191 & 0.999172 & 0.000019  & 0.001864\\
\hline
\end{tabular}
\end{table}

\subsection{PMNS Matrix}

The reconstructed PMNS matrix is
\[
|U_{\rm PMNS}|_{\rm model}
=
\begin{pmatrix}
0.821873 & 0.549885 & 0.148832\\
0.502910 & 0.577635 & 0.642977\\
0.267592 & 0.603294 & 0.751286
\end{pmatrix}.
\]
The comparison with the reference values is shown in
Table~\ref{tab:pmns-comparison}.

\begin{table}[h!]
\centering
\caption{Comparison of the reconstructed PMNS matrix with reference values.}
\label{tab:pmns-comparison}
\begin{tabular}{crrrr}
\hline
Entry & Model & Reference & Difference & Rel. diff. [\%]\\
\hline
11 & 0.821873 & 0.824000 & -0.002127 & -0.258099\\
12 & 0.549885 & 0.547000 & 0.002885  & 0.527392\\
13 & 0.148832 & 0.145000 & 0.003832  & 2.642860\\
21 & 0.502910 & 0.500000 & 0.002910  & 0.582042\\
22 & 0.577635 & 0.582000 & -0.004365 & -0.749951\\
23 & 0.642977 & 0.641000 & 0.001977  & 0.308349\\
31 & 0.267592 & 0.267000 & 0.000592  & 0.221848\\
32 & 0.603294 & 0.601000 & 0.002294  & 0.381766\\
33 & 0.751286 & 0.753000 & -0.001714 & -0.227663\\
\hline
\end{tabular}
\end{table}

The largest CKM deviation occurs in the small \(13\) and \(31\) entries,
where relative deviations are amplified by the small absolute size of the
entries.  The largest PMNS deviation is in the \(13\) entry.  Overall, the
mixing matrices are reproduced at the percent level, while the charged mass
ratios are reproduced at leading order with few-to-ten percent residuals.

\section{Conclusion}

We have presented a compact numerical construction for charged-fermion mass
ratios and fermion mixing matrices based on fixed generation-dependent
exponents.  The same exponent structure defines the leading mass hierarchy
and the overlap factors entering the CKM and PMNS mixing amplitudes.  The
construction uses the fixed structural inputs \(N_c=3\), \(N_w=2\), the
generation index \(G=1,2,3\), and the angular factors
\[
b_{60}=\cos\frac{\pi}{3},
\qquad
b_{45}^{+}=\cos\frac{\pi}{4}+\sin\frac{\pi}{4}.
\]

The leading exponent structure gives the charged-fermion mass ratios at the
few-to-ten percent level.  The independent mixing sines obtained from the
same construction are
\[
s^q_{12}=0.225313537,\qquad
s^q_{23}=0.040055899,\qquad
s^q_{13}=0.003693947,
\]
for the CKM sector, and
\[
s^\ell_{12}=0.556078168,\qquad
s^\ell_{23}=0.650218338,\qquad
s^\ell_{13}=0.148832147,
\]
for the PMNS sector.  After standard unitary reconstruction, the resulting
CKM and PMNS matrices reproduce the reference moduli at the percent level.

The CKM reconstruction uses the phase
\[
\alpha_q
=
67.62^\circ .
\]
The PMNS modulus comparison shown in this work uses \(\alpha_\ell=0\).  The
full CKM and PMNS matrices are reconstructed from the three independent
mixing sines and the corresponding phase choice; their entries are therefore
not counted as independent input quantities.

The present formulation should be understood as a leading-order numerical
structure.  In particular, the charged-fermion mass ratios still show
residual deviations, and the physical neutrino mass-squared splittings are
not fully determined by the leading expression alone.  A possible extension
is to include additional multiplicative mass-dressing factors,
\[
m_A^{\rm phys}(G)
=
m_A^{(0)}(G)\,F_A(G),
\]
where \(m_A^{(0)}(G)\) denotes the leading mass pattern studied here.  Such
corrections are left for future work and are not included in the leading
mixing-overlap factors used in the CKM and PMNS calculations.

The numerical economy of the construction can be summarized as follows.  The
calculation uses fixed structural inputs,
\[
N_c=3,\qquad
N_w=2,\qquad
G=1,2,3,\qquad
\delta_{G1},\qquad
b_{60}=\cos\frac{\pi}{3},\qquad
b_{45}^{+}=\cos\frac{\pi}{4}+\sin\frac{\pi}{4},
\qquad
\Phi_0=\frac{\pi}{2}.
\]
Sector selectors determine whether \(A=u,d,\nu,e\), and signs such as
\[
\sigma_A=\delta_{Au}+\delta_{A\nu}-\delta_{Ad}-\delta_{Ae}
\]
are derived quantities, not independent parameters.  With no continuous
numerical optimization, the same exponent structure generates thirteen
correlated numerical outputs: six charged-fermion mass ratios, six mixing
sines, and the CKM phase used in the unitary reconstruction.  Explicitly,
these are
\[
\left\{
\frac{m_c}{m_u},\frac{m_t}{m_c},
\frac{m_s}{m_d},\frac{m_b}{m_s},
\frac{m_\mu}{m_e},\frac{m_\tau}{m_\mu}
\right\},
\]
\[
\{s^q_{12},s^q_{23},s^q_{13},
s^\ell_{12},s^\ell_{23},s^\ell_{13}\},
\qquad
\alpha_q .
\]
The mass ratios and mixing sines are not statistically independent, since
both are derived from the same sector exponent functions \(L_A(G)\).  They
should therefore be interpreted as correlated consequences of one common
leading structure.  The full CKM and PMNS matrices displayed above are then
obtained from these quantities by standard unitary reconstruction and are
not counted as additional independent outputs.

\end{document}